\documentclass{iaus}
\usepackage{graphicx}

\title[ALFALFA Detects 500 kpc HI Tail] 
{A 500 kpc HI Tail of the Virgo Pair NGC4532/DDO137 Detected by
ALFALFA}

\author[Rebecca A. Koopmann]   
{Rebecca A. Koopmann$^{1,2}$%
 }

\affiliation{$^1$ Department of Physics and Astronomy, Union College,
  Schenectady, NY 12308, \break email: koopmanr@union.edu \\
$^2$ National Astronomy and Ionosphere Center\thanks{
The National Astronomy and Ionosphere Center 
is operated by Cornell University under
a cooperative agreement with the National Science Foundation.}, Space Sciences
   Building, Cornell University, Ithaca NY 14853}

\pubyear{2007}
\volume{244}  
\pagerange{xxx-xxx}
\date{?? and in revised form ??}
\jname{Proceedings Title IAU Symposium}
\editors{Jonathan I. Davies \& Michael D. Disney, eds.}

\def\etal{{\it et al.}}

\def\msun{$M_\odot$}

\begin{document}

\maketitle

\begin{abstract}
HI observations of the Virgo Cluster pair NGC 4532/DDO 137, conducted
as part of the Arecibo Legacy Fast ALFA Survey, reveal an HI feature
extending $\sim$500kpc to the southwest. The structure has a total mass
of up to 7 x 10$^8$ \msun, equivalent to 10\% of the pair HI
mass. Optical R imaging reveals no counterparts to a level of 26.5 mag
arcsec$^{-2}$. The structure is likely the result of galaxy harassment.
\end{abstract}

\firstsection 
\section{Summary and Results}

Cluster environmental interactions (see Boselli \& Gavazzi 2006 for a
review) can produce tails of gas and stars.  The Arecibo Legacy Fast
ALFA (ALFALFA) Survey, a sensitive blind survey of the Arecibo sky
(Giovanelli \etal~2005 and these proceedings), has revealed several HI
clouds without optical components (Kent \etal~2007 and these
proceedings) and a 250 kpc tidal arc emerging from the Sc galaxy NGC
4254 (Haynes \etal~2007 and Giovanelli in these proceedings).  ALFALFA
has recently detected an even larger tidal feature associated with the
Virgo Cluster Sm pair NGC 4532/DDO 137.  This system was already known
to be peculiar: both galaxies have extended HI disks and share a
common HI envelope extended over 150 kpc (Hoffman et al. 1993, 1999).

\begin{figure}
\centering
\includegraphics[scale=0.57]{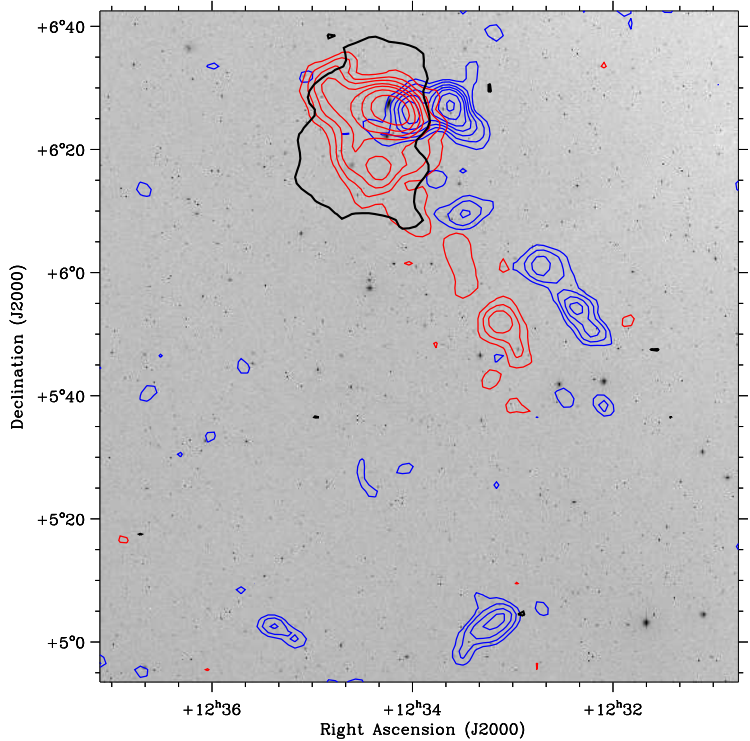}
\caption{ALFALFA HI flux contours superposed on a DSS2 Blue Image. The
black contour at 0.4 Jy beam$^{-1}$ km s$^{-1}$, integrated over
1956 - 2139 km s$^{-1}$, encompasses the approximate area of the HI envelope 
detected by Hoffman \etal (1993). Blue contours 
show tail emission integrated over 1784 - 1836 km s$^{-1}$,
with contours at 0.17, 0.25, 0.35, 0.45, 0.55, 0.7, 0.9 Jy beam$^{-1}$ km s$^{-1}$.
Red contours show tail emission 
integrated over 1868 and 1930 km$^{-1}$, with contours at 
0.23, 0.4, 0.6, 0.9, 1.2, 1.7, and 2.5 Jy beam$^{-1}$ km s$^{-1}$.
\label{fig1}}
\end{figure}

ALFALFA observations of the NGC 4532/DDO 137 tail structure are shown
in Figure~\ref{fig1}.  The HI envelope containing and within the
immediate vicinity of the pair (black contour) has an HI mass of
6.2 x 10$^9$ \msun, consistent with that of Hoffman \etal~(1999).  All
of the emission in the tail is blueshifted with respect to the pair HI
envelope. The total mass contained within discrete clumps in the tail
is 4.0 x 10$^8$ \msun.  The total mass of the tail, including an upper
limit for emission below the ALFALFA limiting column density, is
$\sim$7 x 10$^8$ \msun, or $\sim$10\% of the pair HI mass.  R imaging
of the tail system was carried out at Wise Observatory and the WIYN
0.9-m telescope in May 2007. No optical counterparts for the main HI
clumps have been found to a limiting R magnitude of 26.5 mag
arcsec$^{-2}$.  Further results are presented in Koopmann \etal~(2007,
in preparation.)  VLA syntheis observations of the system are planned.

The observations of
NGC 4532/DDO 137 are consistent with several predictions from the
simulations of high speed hyperbolic encounters between cluster
galaxies by Bekki, Koribalski \& Kilborn (2005) and Duc (these
proceedings):
(a) gas tails that are stretched over several hundred kpc,
(b) double tails or tails that span a large spatial area,
(c) formation of relatively isolated clumps within the tail,
and (d) stellar tails fainter than 30 mag arcsec$^{-2}$.

The discovery of an extended tail with discrete clumps so distant from
the parent galaxy suggests a tidal explanation for at least some
isolated HI clouds with no optical counterparts, such as those
recently discovered in the Virgo Cluster (Kent \etal~ 2007 and
these proceedings). 

\begin{acknowledgments}

The author is grateful for partial support from NAIC, travel support
      from the Mellon Foundation, and for the hospitality of the
      Cornell University Astronomy Department during a sabbatic visit.

\end{acknowledgments}

\end{document}